\UseRawInputEncoding
\documentclass[%
reprint,
superscriptaddress,
nofootinbib,
 amsmath,amssymb,
 aps,
]{revtex4-1}
\usepackage{graphicx}
\usepackage{subfigure}
\usepackage{dcolumn}
\usepackage{bm}
\usepackage[%
bookmarksnumbered,
bookmarksopen,
colorlinks,
citecolor=blue,
linkcolor=blue,
]{hyperref} 

\def\esym{$E_{sym}(\rho)$~}
\def\rpi {$\pi^-/\pi^+$~}
\def\es0{$E_{sym}(\rho_0)$~}
\def\us0{$U_{sym}^{\infty}(\rho_{0})$}
\begin{document}


\title{Effects of incompressibility $K_{0}$ in heavy-ion collisions at intermediate energies}

\author{Xiao-Xiao Long}
\affiliation{School of Physics and Electronic Science, Guizhou Normal University, Guiyang 550025, China}
\author{Gao-Feng Wei}\email{Corresponding author: wei.gaofeng@gznu.edu.cn}
\affiliation{School of Physics and Electronic Science, Guizhou Normal University, Guiyang 550025, China}
\affiliation{Guizhou Provincial Key Laboratory of Radio Astronomy and Data Processing, Guizhou Normal University, Guiyang 550025, China}

\begin{abstract}

Within the possible least uncertainty on the nuclear incompressibility $K_{0}$, we examine effects of $K_{0}$ in heavy-ion collisions at intermediate energies. Based on simulations of Au + Au collision at 400 MeV/nucleon using an isospin- and momentum-dependent transport model, we find that the incompressibility $K_{0}$ indeed affects significantly the attainable density in central regions, and thus the particle productions and/or distributions at final states, e.g., nucleon rapidity distributions and yields of charged pions. Nevertheless, through examining the free neutron over proton ratios $n/p$, the neutron-proton differential transverse and directed flows as well as the charged pion ratio $\pi^{-}/\pi^{+}$ and its kinetic energy distribution, we find that these observables are less affected by the uncertainty of $K_{0}$, but mainly sensitive to the slope of symmetry energy at the saturation density. We also compare and discuss our results with the corresponding data.
\end{abstract}

\maketitle


\section{introduction}\label{introduction}

As one of central issues in isospin nuclear physics, the symmetry energy \esym at suprasaturation densities has been a long-standing interest due to its importance in understanding the properties of radioactive nuclei and evolution of supernova and neutron stars~\cite{Ste05,Lat12,Hor14,Heb15,Bal16,Oer17,Tew18}. In terrestrial laboratories, heavy-ion collisions (HICs) with rare isotopes provide a unique opportunity to generate directly the isospin asymmetric nuclear matter at high densities, and thus enable one to extract the information about the \esym at high densities through comparing the theoretical simulations of isospin observables with the corresponding data~\cite{Bar05,Kol05,li08,Lyn09}.

The incompressibility $K_{0}$ of nuclear matter at the saturation density $\rho_{0}$, as an important input of most microscopic and/or phenomenological heavy-ion transport models, affects the attainable density in collision regions and thus the particle productions and/or distributions at final states. Naturally, the accuracy of $K_{0}$ affects the quantitative extraction of the symmetry energy using HIC models. However, the possible tightest constraint or least uncertainty one currently obtains on the $K_{0}$ is $230\pm30$ MeV~\cite{Sturm01,Fuch01,Dan02,Hart06,Pan93,Zhang94}. This uncertainty on the value of $K_{0}$ naturally prevents one from quantitatively extracting the high density symmetry energy information using isospin observables. Actually, some literatures have already involved in investigations of the uncertainty of $K_{0}$ in HICs and other aspects. For example, using a T\"{u}bingen quantum molecular dynamics model, Ref.~\cite{Cozma18} discussed effects of $K_{0}$ within a range from 210 to 280 MeV on the nucleon elliptic flows, and Ref.~\cite{Zan12} examined effects of $K_{0}$ within a range from 195 to 225 MeV on the nucleus-nucleus dynamic potential in fusion reactions, and Ref.~\cite{Bon18} studied effects of $K_{0}$ within a range from 210 to 240 MeV on the giant resonances in $^{40,48}$Ca, $^{68}$Ni, $^{90}$Zr, $^{116}$Sn, $^{144}$Sm and $^{208}$Pb. Nevertheless, a systematical study related to the effects of $K_{0}$ on the symmetry energy observables is rarely reported. On the other hand, it is well known that central heavy-ion reactions at intermediate energies play a special role in determination of the symmetry energy especially above twice times saturation density. Therefore, it is naturally necessary to study effects of $K_{0}$ in central heavy-ion reactions at intermediate energies. To this end, we perform a central Au + Au collision at 400 MeV/nucleon to study effects of $K_{0}$ within the possible tightest constraint, i.e., $K_{0}=230\pm30$ MeV, on the pion and nucleon observables. It is shown that the $K_{0}$ indeed affects significantly the attainable density in collision regions, and thus the particle productions at final states, e.g., charged pion multiplicities etc. However, we find that the free neutron over proton ratios $n/p$, the neutron-proton differential transverse and directed flows as well as the charged pion ratio $\pi^{-}/\pi^{+}$ and its kinetic energy distribution could reduce significantly the effects of uncertainties of $K_{0}$ and thus show more sensitivities to the high density behavior of symmetry energy.

In the following, we first describe briefly the used isospin- and momentum-dependent Boltzmann-Uehling-Uhlenbeck transport model (IBUU)~\cite{IBUU1,IBUU2} in Sec.~\ref{Model}. We then discuss our results in Sec.~\ref{Results and Discussions}. A summary will be given in Sec.~\ref{Summary}.
\begin{figure}[hbt]
	\includegraphics[width=\columnwidth]{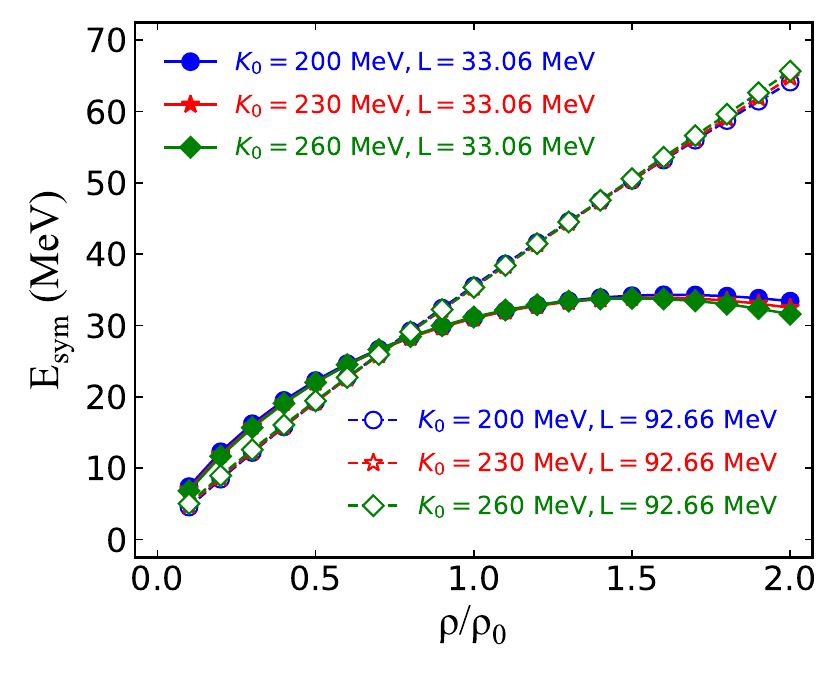}
	\caption{(Color online)Density dependence of the $E_{sym}(\rho)$ with different $K_{0}$ and $L$.} 
	\label{sym}
\end{figure}
\section{The Model}\label{Model}

The present study is carried out within an updated version of IBUU transport model~\cite{Wei23}. In this version, we adopt a separate density-dependent scenario~\cite{Wei20} for a more delicate treatment of the in-medium many-body force effects as in Refs.~\cite{Xu10,Chen14}. Also, we introduce a parameter $z$~\cite{Wei22} as in Ref.~\cite{Xu15} to mimic the value of \esym at $\rho_{0}$ and $\tilde{\rho}\approx2\rho_{0}/3$ to meet the best knowledge of \esym at the two densities one has obtained so far, e.g., $E_{sym}(\rho_{0})=32.5\pm2$ MeV~\cite{Cozma18}, $32.5\pm3.2$ MeV~\cite{Wang18}, $33.0^{+2}_{-1.8}$ MeV~\cite{Ess21}, $38.3\pm4.7$~MeV~\cite{Ree21}, $35.3\pm2.8$ MeV~\cite{Est21}, and $E_{sym}(\tilde{\rho}=0.1~{\rm fm}^{-3})=25.5\pm1$~MeV~\cite{Brown13}, $E_{sym}(\tilde{\rho}=0.11~{\rm fm}^{-3})=26.2\pm1$ MeV~\cite{Wang13}, $E_{sym}(\tilde{\rho}=0.11~{\rm fm}^{-3})=26.65\pm0.2$ MeV~\cite{Zhang13}. As to the high density behavior of \esym, we use the parameter $x$ to control the slope value $L\equiv{3\rho({dE_{sym}}/d\rho})$ of \esym at $\rho_{0}$ as in the original IBUU model~\cite{IBUU1,IBUU2}. All of these features have been incorporated into the present used model, see Ref.~\cite{Wei23} for details. Specifically, the isospin and momentum dependent nuclear interaction (MDI) used is expressed as:
\begin{eqnarray}
U(\rho,\delta ,\vec{p},\tau ) &=&A_{u}\frac{\rho _{-\tau }}{\rho _{0}}%
+A_{l}\frac{\rho _{\tau }}{\rho _{0}}+\frac{B}{2}{\Big(}\frac{2\rho_{\tau} }{\rho _{0}}{\Big)}^{\sigma }(1-x)  \notag \\
&+&\frac{2B}{%
	\sigma +1}{\Big(}\frac{\rho}{\rho _{0}}{\Big)}^{\sigma }(1+x)\frac{\rho_{-\tau}}{\rho}{\big[}1+(\sigma-1)\frac{\rho_{\tau}}{\rho}{\big]}
\notag \\
&+&\frac{2C_{l }}{\rho _{0}}\int d^{3}p^{\prime }\frac{f_{\tau }(%
	\vec{p}^{\prime })}{1+(\vec{p}-\vec{p}^{\prime })^{2}/\Lambda ^{2}}
\notag \\
&+&\frac{2C_{u }}{\rho _{0}}\int d^{3}p^{\prime }\frac{f_{-\tau }(%
	\vec{p}^{\prime })}{1+(\vec{p}-\vec{p}^{\prime })^{2}/\Lambda ^{2}},
\label{IMDIU}
\end{eqnarray}%
where $\tau=1$ for neutrons and $-1$ for protons, and $A_{u}$, $A_{l}$, $C_{u}(\equiv C_{\tau,-\tau})$ and $C_{l}(\equiv C_{\tau,\tau})$ are expressed as
\begin{eqnarray*}
A_{l}&=&A_{l0}+U_{sym}^{\infty}(\rho_{0}) - \frac{2B}{\sigma+1}\notag \\
&\times&\Big{[}\frac{(1-x)}{4}\sigma(\sigma+1)-\frac{1+x}{2}\Big{]},  \\
A_{u}&=&A_{u0}-U_{sym}^{\infty}(\rho_{0}) + \frac{2B}{\sigma+1}\notag \\
&\times&\Big{[}\frac{(1-x)}{4}\sigma(\sigma+1)-\frac{1+x}{2}\Big{]},\\
C_{l}&=&C_{l0}-2\big{(}U_{sym}^{\infty}(\rho_{0})-2z\big{)}\frac{p_{f0}^{2}}{\Lambda^{2}\ln \big{[}(4p_{f0}^{2}+\Lambda^{2})/\Lambda^{2}\big{]}},\\
C_{u}&=&C_{u0}+2\big{(}U_{sym}^{\infty}(\rho_{0})-2z\big{)}\frac{p_{f0}^{2}}{\Lambda^{2}\ln \big{[}(4p_{f0}^{2}+\Lambda^{2})/\Lambda^{2}\big{]}}.
\end{eqnarray*}
The eight parameters embedded in above expressions, i.e., $A_{l0}$, $A_{u0}$, $B$, $\sigma$, $C_{l0}$, $C_{u0}$, $\Lambda$ and $z$, are determined by fitting eight experimental and/or empirical constraints on properties of nuclear matter at $\rho_{0}=0.16$~fm$^{-3}$.  Among them, the first seven are the binding energy $-16$~MeV, the pressure $P_{0}=0$~MeV/fm$^{3}$ , the isoscalar effective mass $m^{*}_{s}=0.7m$, the isoscalar potential at infinitely large nucleon momentum $U^{\infty}_{0}(\rho_{0})=75$~MeV, the isovector potential at infinitely large nucleon momentum $U^{\infty}_{sym}(\rho_{0})=-100$~MeV, and the \esym at $\rho_{0}$ and $\tilde{\rho}\approx2\rho_{0}/3$. The  eighth is $K_{0}$ that we are going to examine. To this end, we take three values for $K_{0}$ within the possible least uncertainty as aforementioned in this study, i.e., 200, 230 and 260~MeV. For these different settings of $K_{0}$, the corresponding values of $A_{l0}$, $A_{u0}$, $B$ and $\sigma$ are shown in Table~\ref{tableI}, and the values of $C_{l0}$, $C_{u0}$ and $\Lambda$ are $C_{l0}=-60.486$ MeV, $C_{u0}=-99.702$ MeV and $\Lambda =2.424p_{f0}$ , where $p_{f0}$ refers to the nucleon Fermi momentum in symmetric nuclear matter (SNM) at $\rho_{0}$. Also, to make the symmetry energy observable more clearly reflecting effects of $K_{0}$, we adjust the values of $x$ and $z$ for different $K_{0}$ settings to ensure the identical slope $L$ of \esym at $\rho_{0}$ as shown in Fig.~\ref{sym}. In addition, for a certain $K_{0}$, we also take four different settings for $L$, and to enable us to compare effects of $K_{0}$ and $L$ on the symmetry energy observables. It should be mentioned that the parameter $z$ should ensure the values of symmetry energy at both $\rho_{0}$ and $\tilde{\rho}\approx2\rho_{0}/3$ to be basically within the allowed ranges as indicated in Refs.~\cite{Cozma18,Wang18,Ess21,Ree21,Est21,Brown13,Wang13,Zhang13} as aforementioned. In this study, we limit the value of symmetry energy at $\tilde{\rho}\approx2\rho_{0}/3$ within the range of $25.5\pm1$~MeV, while that at $\rho_{0}$ is determined as $E_{sym}(\rho_{0})=32.5+z$~MeV. The specific values of $x$ and $z$ and the corresponding $L$ as well as the $E_{sym}(2\rho_{0}/3)$ are shown in Table~\ref{tableII}.
\begin{table}[hbt]
	\caption{\label{tableI}
		The values of $A_{l0}$, $A_{u0}$, $B$, $\sigma$ and the resulting $K_{0}$.}
	\begin{ruledtabular}
		\begin{tabular}{cccc}
			\textrm{$A_{l0}$=$A_{u0}$ (MeV)}&
			\textrm{$B$ (MeV)}&
			\textrm{$\sigma$}&
			\textrm{${K_{0}}$ (MeV)}\\
			\colrule
			$-455.726$    & $530.726$    & $1.0646$  & $200$\\
			$-66.963$     & $141.963$    & $1.2652$  & $230$\\
			$-12.992$     & $87.992$     & $1.4657$  & $260$\\
		\end{tabular}
	\end{ruledtabular}
\end{table}
\begin{table}[hbt]
	\caption{\label{tableII}
		The values of $x$, $z$ and the resulting $K_{0}$, $L$ and $E_{sym}(2\rho_{0}/3)$.}
	\begin{ruledtabular}
		\begin{tabular}{ccccccccccccc}
			\textrm{${K_{0}}$ (MeV)}&
			\textrm{$x$}&
			\textrm{$z$ (MeV)}&
			\textrm{$L$ (MeV)}&
			\textrm{${E_{sym}(2\rho_{0}/3)}$ (MeV)}\\
			\colrule
			$200$    & $0.54$     & $-1.408$    & $33.06$    & $26.06$\\
			$200$    & $0.08$     & $0.474$     & $61.95$    & $25.24$\\
			$200$    & $-0.38$    & $3.066$     & $92.66$    & $24.93$\\
			$200$    & $-0.84$    & $5.8$       & $123.73$   & $24.71$\\
			\hline
			$230$    & $0.6$     & $-1.482$    & $33.06$     & $25.90$\\
			$230$    & $0.2$     & $0.326$     & $61.95$     & $25.10$\\
			$230$    & $-0.2$    & $2.844$     & $92.66$     & $24.80$\\
			$230$    & $-0.6$    & $5.505$     & $123.73$    & $24.60$\\
			\hline
			$260$    & $0.66$     & $-1.3$      & $33.06$      & $25.98$\\
			$260$    & $0.31$     & $0.418$     & $61.95$      & $25.18$\\
			$260$    & $-0.04$    & $2.845$     & $92.66$      & $24.88$\\
			$260$    & $-0.39$    & $5.415$     & $123.73$     & $24.67$\\
		\end{tabular}
	\end{ruledtabular}
\end{table}
\begin{figure}[thb]
	\includegraphics[width=\columnwidth]{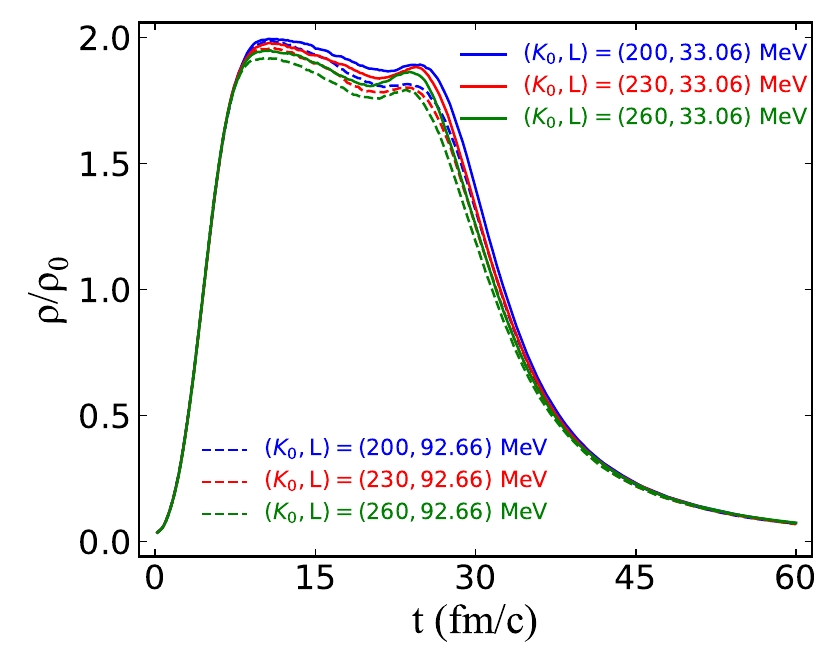}
	\caption{(Color online)Evolution of reduced average densities $\rho/\rho_0$ in central spherical regions with a radius of 2 fm in Au+Au collisions at 400 AMeV with different $K_{0}$ and $L$.}
	\label{den}
\end{figure}
\section{Results and Discussions}\label{Results and Discussions}

Now, we present the results of Au+Au collisions at 400MeV/nucleon with an impact parameter of $b=0-2$ fm, corresponding to a typical reaction with a reduced impact parameter of $b_{0}\le0.15$ at 400MeV/nucleon carried out at the FOPI detector~\cite{FOPI07,FOPI10,FOPI12}.
\begin{figure}[thb]
	\includegraphics[width=\columnwidth]{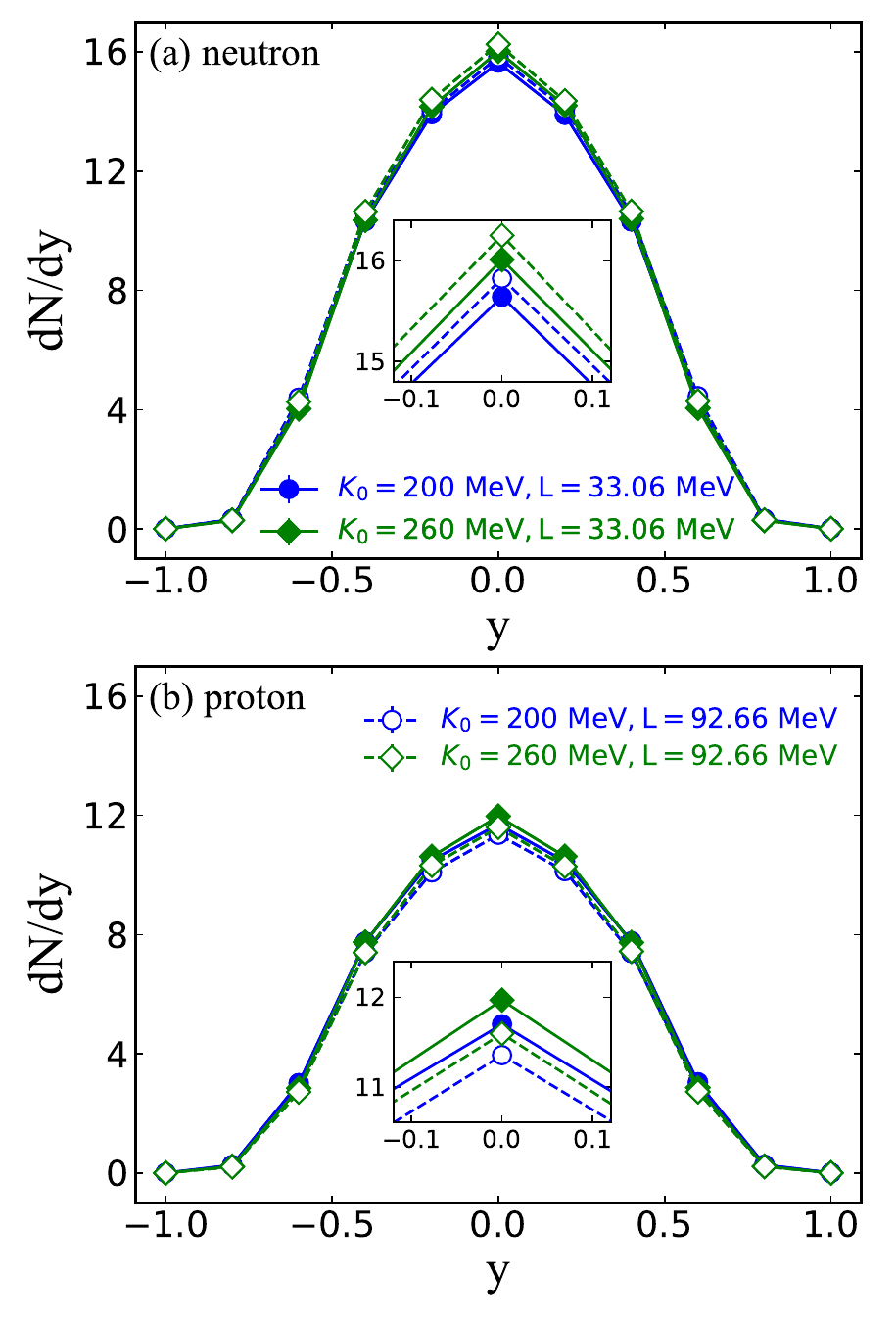}
	\caption{(Color online)Rapidity distributions of neutrons (a) and protons (b) in Au+Au collisions at 400 AMeV with different combinations of $K_{0}$ and $L$.}
	\label{np}
\end{figure}
\begin{figure}[thb]
	\includegraphics[width=\columnwidth]{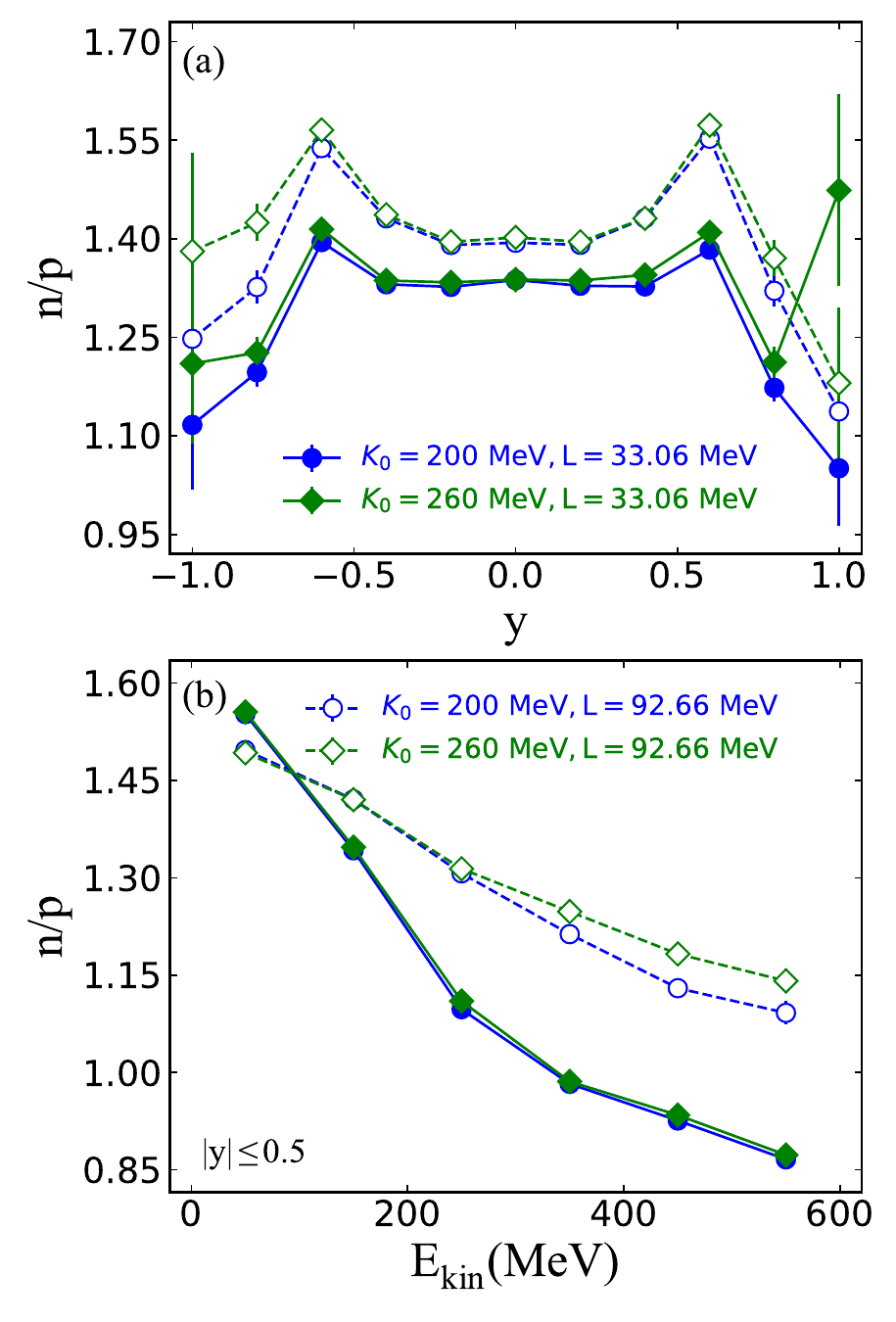}
	\caption{(Color online) Rapidity (a) and kinetic energy (b) distributions of $n/p$ ratios in Au+Au collisions at 400 AMeV with different combinations of $K_{0}$ and $L$.}
	\label{npr}
\end{figure}
\subsection{Nucleon observables}\label{Nucleon observables}
Shown in Fig.~\ref{den} are the evolutions of reduced average densities in central spherical regions with a radius of 2 fm. To study how the $K_{0}$ and $L$ affect the attainable compression densities in the reactions, we take 6 different combinations of $K_{0}$ and $L$ in simulations of Au + Au collisions. It is seen that, for a certain $L$, the compression density reached in central regions is significantly larger with a smaller value of $K_{0}$ than that with a larger one; while for a certain $K_{0}$, the compression density reached is significantly larger with a soft symmetry energy (i.e., a smaller slope value $L$) than that with a stiff one (i.e., a larger slope value $L$). These observations are exactly the features belonging to HICs at intermediate energies~\cite{Bar05,li08} since the effect of symmetry energy on the compression density is decreasing even to a negligible degree as the beam energy increases up to approximate 1 GeV and above~\cite{Wei23}. However, if one compares semiquantitatively effects between $K_{0}$ and $L$ on the compression density, e.g., varying $K_{0}$ from 260 to 200~MeV and $L$ from 92.66 to 33.06~MeV, their relative changes are $(260-200)/[(260+200)/2]\times\%\approx26\%$ and $(92.66-33.06)/[(92.66+33.06)/2]\times\%\approx94.8\%$, one can explicitly find from Fig.~\ref{den} that the effect of $K_{0}$ on the compression density is obviously dominant than that of $L$, reflecting the fact that the nuclear compression is overall dominated by the bulk EoS of nuclear matter. Naturally, one expects that these features could be reflected by the nucleon observables at final states. To this end, we show in Fig.~\ref{np} the rapidity distributions of free neutrons and protons at final states, where the criterion of free nucleons is defined as the relative distances $\Delta_{R}>3.575$~fm or momenta $\Delta_{p}>0.3$~GeV/c as in coalescence models~\cite{Gyul83,Aich87,Koch90,Nagl96,Sche99}. It is seen that, for a certain $L$, both the free neutrons and protons are more with a larger $K_{0}$ than those with a smaller $K_{0}$ as shown in insets, since the isoscalar potentials have approximately identical effects on neutrons and protons. While for a certain $K_{0}$, it is obvious to see that the variation tendency of neutrons is completely opposite to that of protons when varying the $L$, reflecting the fact that the symmetry potential/energy has opposite effects on neutrons and protons at high densities, i.e., repulsion on neutrons but attraction on protons. To these observations, one naturally expects the ratios of free neutrons over protons could reduce the isoscalar potential effects and enlarge the symmetry potential/energy effects.

\begin{figure}[thb]
	\includegraphics[width=\columnwidth]{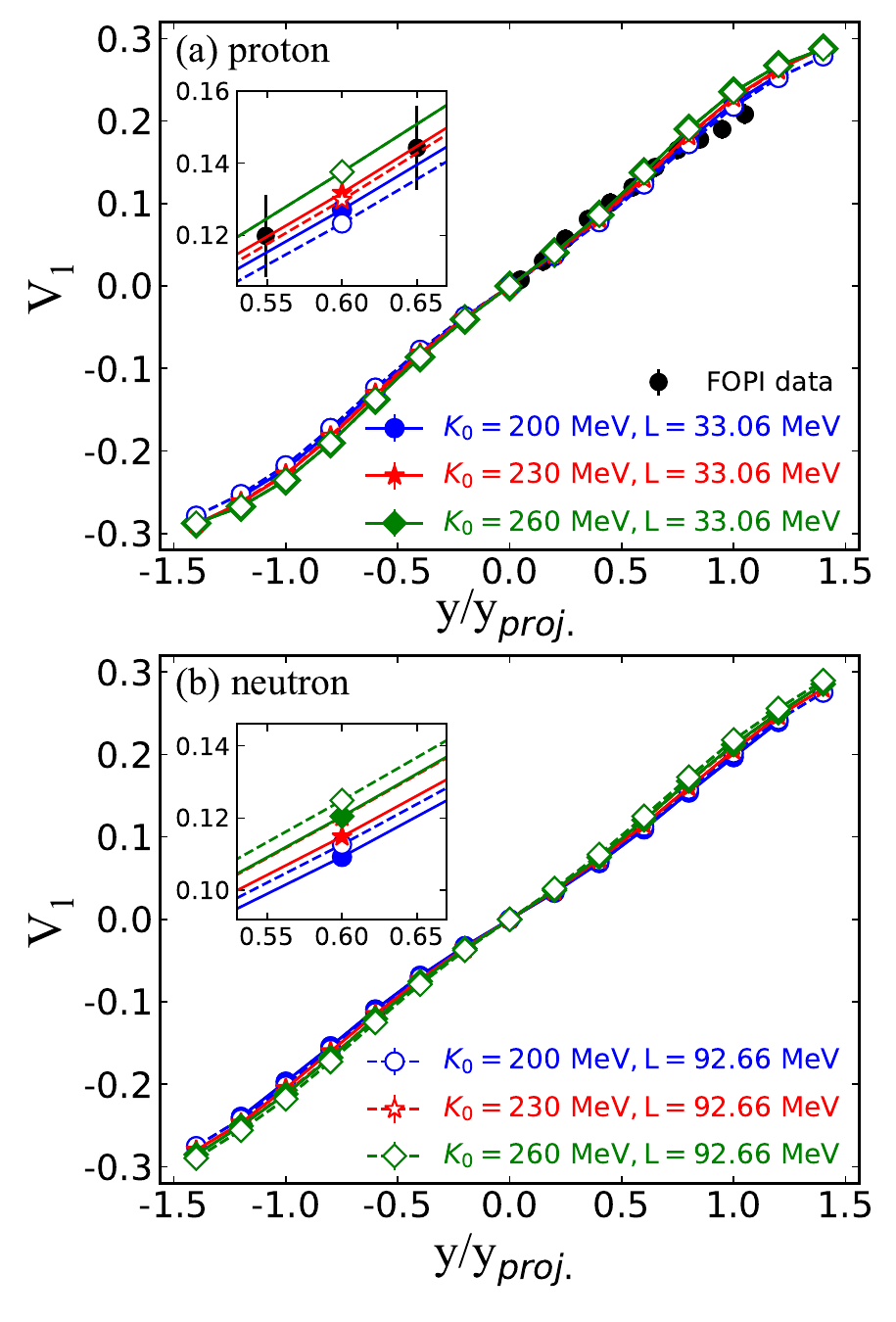}
	\caption{(Color online)Directed flows of protons (a) and neutrons (b) in Au+Au collisions at 400MeV/nucleon with different combinations of $K_{0}$ and $L$.}
	\label{v1}
\end{figure}
\begin{figure}[thb]
	\includegraphics[width=\columnwidth]{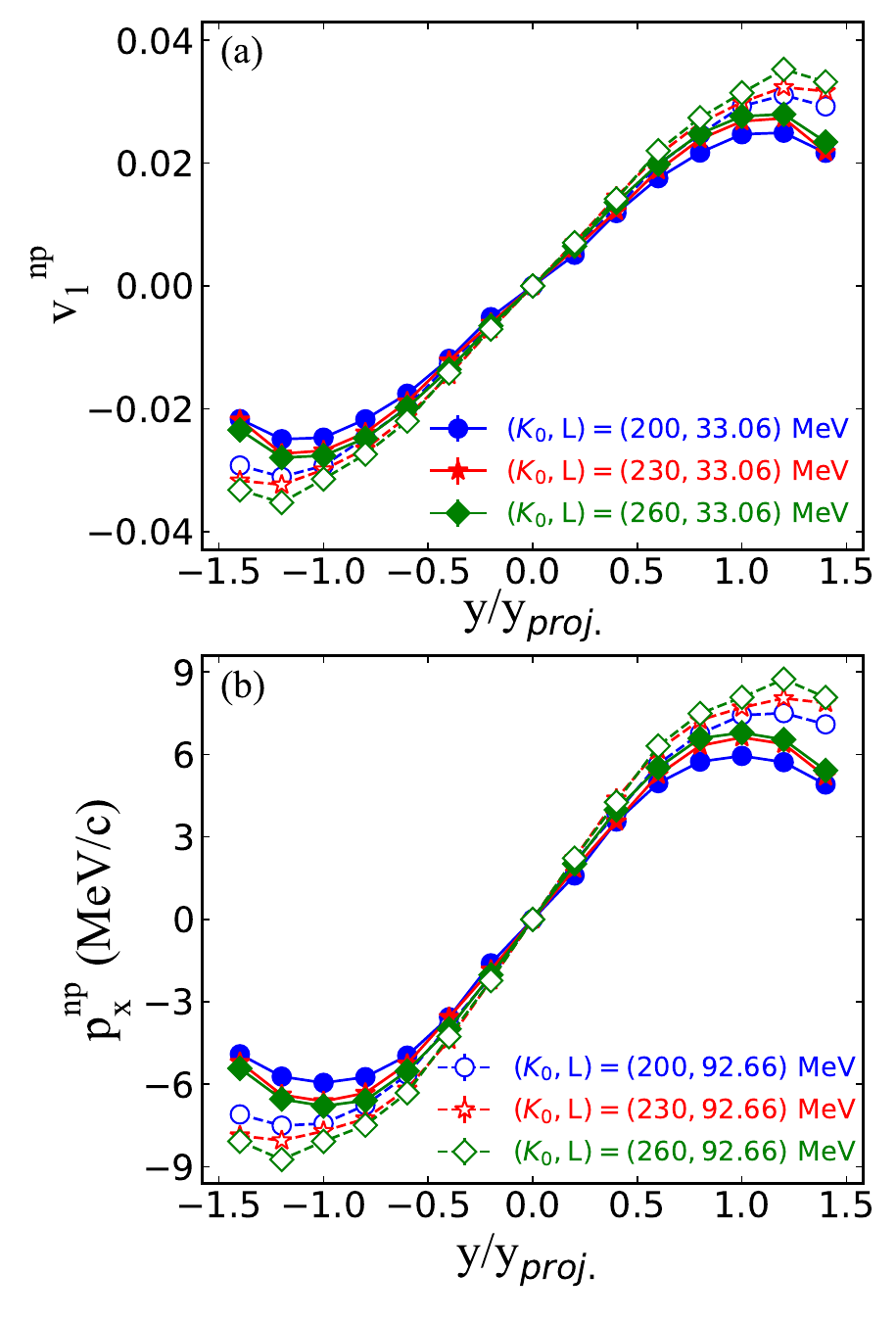}
	\caption{(Color online)Free neutron-proton differential directed (a) and transverse (b) flows in Au+Au collisions at 400MeV/nucleon with different combinations of $K_{0}$ and $L$.}
	\label{v1np}
\end{figure}

Shown in Fig.~\ref{npr}(a) are the rapidity distributions of neutrons over protons $n/p$ with different combinations of $K_{0}$ and $L$. As one expected, the $n/p$ ratios indeed reduce significantly effects of $K_{0}$ and enlarge those of $L$. Moreover, because nucleons at midrapidities are mainly from early emissions during the compression stage, and thus carry the information of symmetry energy at high densities. Therefore, we can observe a larger $n/p$ ratio with a stiff symmetry energy than that with a soft one. Similarly, it is seen that the kinetic energy distributions of $n/p$ ratios at midrapidities are mainly sensitive to the high density behavior of symmetry energy as shown in Fig.~\ref{npr}(b).
\subsection{Flow observables}\label{Flow observables}
Collective motions of final state nucleons are the direct reflections of the pressure created in HICs and thus are closely related to the equation of state of dense nuclear matter. Therefore, in this subsection, we examine how the $K_{0}$ and $L$ affect the collective motions of final state nucleons. In our studied reactions,  the main collective motions could be reflected by the directed flows $v_{1}$ and/or transverse flows $p_{x}$. Shown in Fig.~\ref{v1} are the rapidity dependent directed flows of free neutrons and protons at final states. The insets are local amplification to explicitly show effects of $K_{0}$ and $L$. To compare with the corresponding FOPI data~\cite{FOPI07,FOPI10,FOPI12}, we use the same reduced rapidity as in Refs.~\cite{FOPI07,FOPI10,FOPI12}, i.e., $y/y_{proj.}$. First, it is seen that our results are consistent with the data. Second, similar to the observations in rapidity distributions of neutrons and protons shown in Fig.~\ref{np}, the effects of symmetry energy/potential are completely opposite for neutrons and protons, while the effects of $K_{0}$ on neutrons and protons are approximately identical. Therefore, we turn to the free neutron-proton differential directed and transverse flows defined as~\cite{Bar05,Wei22,li00,li02},
\begin{eqnarray}
v_{1}^{np} &=&\frac{N_{n}(y)}{N(y)}{ \left\langle v_{1}^{n}(y)\right\rangle}-\frac{N_{p}(y)}{N(y)}{ \left\langle v_{1}^{p}(y)\right\rangle},
\end{eqnarray}
\begin{eqnarray}
p_{x}^{np} &=&\frac{N_{n}(y)}{N(y)}{ \left\langle p_{x}^{n}(y)\right\rangle}-\frac{N_{p}(y)}{N(y)}{ \left\langle p_{x}^{p}(y)\right\rangle},
\end{eqnarray}
where $N_{n}(y)$, $N_{p}(y)$, and $N(y)$ respectively represents the total number of free neutrons, protons, and nucleons at rapidities $y$. 
Shown in Fig.~\ref{v1np} are the corresponding simulations of free neutron-proton differential directed $v_{1}^{np}$ and transverse $p_{x}^{np}$ flows. One can see that both the $v_{1}^{np}$ and $p_{x}^{np}$ indeed reduce significantly effects of $K_{0}$ and show more sensitivities to the symmetry energy/potential. 
\begin{figure}[thb]
	\includegraphics[width=\columnwidth]{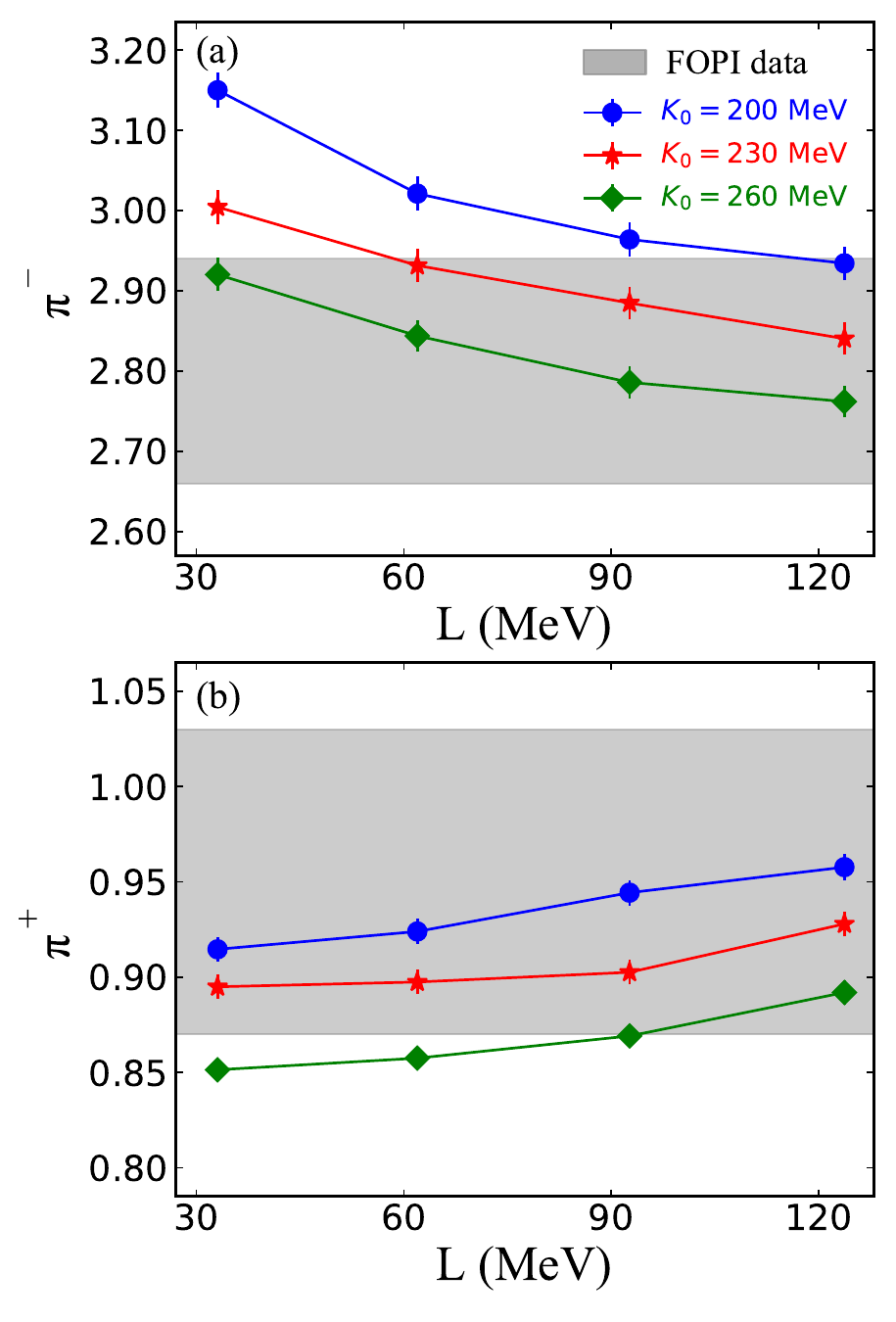}
	\caption{(Color online)Multiplicities of $\pi^{-}$ (a) and $\pi^{+}$ (b) in Au+Au collisions at 400MeV/nucleon with different combinations of $K_{0}$ and $L$.}
	\label{pion}
\end{figure}
\begin{figure}[thb]
	\includegraphics[width=\columnwidth]{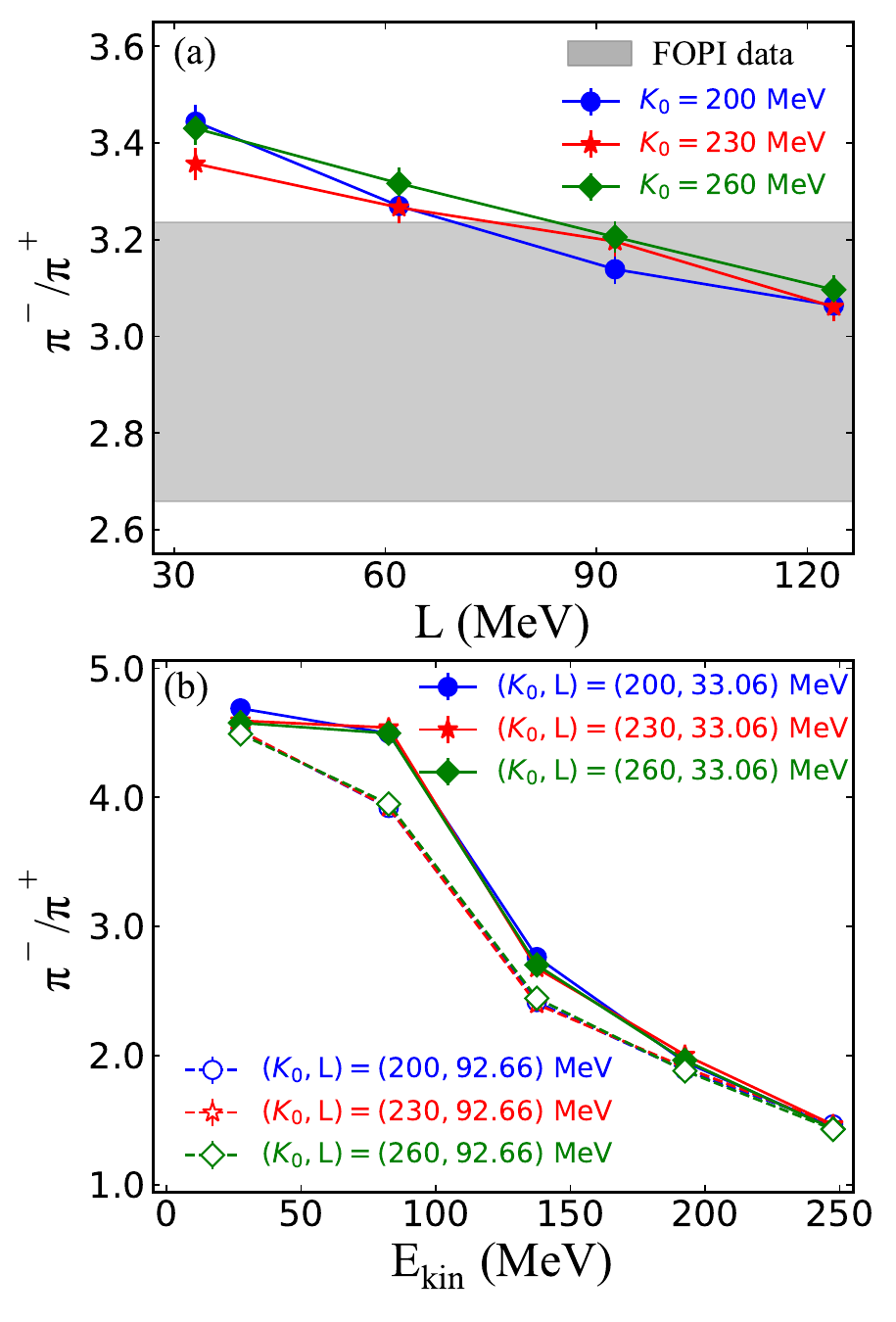}
	\caption{(Color online)Ratios $\pi^{-}/\pi^{+}$ as a function of $L$ (a) and pion kinetic energy (b) in Au+Au collisions at 400MeV/nucleon with different combinations of $K_{0}$ and $L$.}
	\label{pionr}
\end{figure}
\subsection{Pion observables}\label{Pion observables}
In HICs at intermediate energies, pions are produced mostly from the decay of $􏰍\Delta(1232)$ resonances. Specifically, for the charged pions, the main channels are $\Delta^{-}\leftrightarrow n+\pi^{-}$ and $\Delta^{++}\leftrightarrow p+\pi^{+}$. On the other hand, the main channels of producing $\Delta^{-}$ and $\Delta^{++}$ from nucleon-nucleon collisions at high densities are $nn\rightarrow p+\Delta^{-}$ and $pp\rightarrow n+\Delta^{++}$. Equivalently, one can view the production of $\pi^{-}$ as mainly from inelastic $nn\rightarrow pn\pi^{-}$ channels while that of $\pi^{+}$ mainly from $pp\rightarrow pn\pi^{+}$ channels. This is why the ratio \rpi is sensitive to the high density behavior of nuclear symmetry energy in HICs.

Shown in Fig.~\ref{pion} are the multiplicities of $\pi^{-}$ and $\pi^{+}$ in Au+Au collisions at 400MeV/nucleon with different combinations of $K_{0}$ and $L$. First, it can be observed that, consistent with the previous observations using most transport models, the yields of $\pi^{-}$ and $\pi^{+}$ are sensitive to $L$, and the sensitivity of $\pi^{-}$ is greater than that of $\pi^{+}$. Moreover, the variation tendency of $\pi^{-}$ with $L$ is opposite to that of $\pi^{+}$ similar to that of nucleons. At the same time, we can find that yields of both $\pi^{-}$ and $\pi^{+}$ are also sensitive to the $K_{0}$. More specifically, both $\pi^{-}$ and $\pi^{+}$ are more produced in collisions with a smaller $K_{0}$ due to a larger compression formed in collisions as shown in Fig.~\ref{den}. To this observation, one naturally expects that the ratio \rpi could reduce significantly effects of $K_{0}$ and thus show more sensitivities to the high density behavior of symmetry energy. Indeed, this can be demonstrated by the total \rpi ratios and the kinetic energy distributions as shown in Fig.~\ref{pionr}. In addition, we also show the corresponding data in Figs.~\ref{pion} and \ref{pionr}. It can be found that extracting the information of high density symmetry energy from multiplicities of both $\pi^{-}$ and $\pi^{+}$ depends seriously on the used $K_{0}$, while that from both total \rpi ratio and its kinetic energy distributions is free of uncertainties of $K_{0}$. 

Before ending this part, we give a useful remark on the kinetic energy distributions of \rpi ratios. We note that our \rpi ratio at high kinetic energies is insensitive to the symmetry energy, 
while the study on subthreshold pion production from the pBUU~\cite{Hong14} model appears that the \rpi ratio at high kinetic energies is still and even more sensitive to the symmetry energy, and a soft symmetry energy corresponds to a large \rpi ratio. Moreover, we find from the spectral pion ratio simulated with the dcQMD model~\cite{Est21} that the sensitivity of \rpi ratio to the symmetry energy will cross as the transverse momenta of pions increase, i.e., for pions with the low transverse momenta, the \rpi ratio is large with a soft symmetry energy, but for pions with the high transverse momenta, the opposite is true. It appears the sensitivity of \rpi ratio at high kinetic energies or high transverse momenta is still uncertain and needs to be further studied. The discrepancies of \rpi ratios at high kinetic energies and/or transverse momenta might originate from the different $\Delta$ potential that affects the decay of $\Delta$ and thus the attainable kinetic energy for pions. Another possibility is the pion potential that affects the propagation of pions in nuclear medium and thus the kinetic energy distributions of pions at final states. Therefore, it will be interesting to see how these factors affect the kinetic energy distributions of \rpi ratios, especially at high kinetic energies.
\section{Summary}\label{Summary}
In summary, we have studied the effects of incompressibility $K_{0}$ within its possible least uncertainty in HICs at intermediate energies. 
It is shown that the $K_{0}$ indeed affects significantly the attainable density in central collision regions, and thus the particle productions and/or distributions at final states, e.g., nucleon rapidity distributions and yields of charged pions. However, considering that the effects of $K_{0}$ on neutrons and protons are approximately identical, we have examined and found that the free neutron over proton ratios $n/p$, the neutron-proton differential transverse $p_{x}^{np}$ and directed $v_{1}^{np}$ flows could reduce significantly effects of $K_{0}$ and thus show more sensitivities to the symmetry energy. Similarly, the \rpi ratio and its kinetic energy distributions are also found to be less affected by the uncertainty of $K_{0}$, but mainly sensitive to the slope of symmetry energy at $\rho_{0}$. 

This work is supported by the National Natural Science Foundation of China under grant Nos.11965008, 11405128 and Guizhou Provincial Science and Technology Foundation under Grant No.[2020]1Y034, and the PhD-funded project of Guizhou Normal university (Grant No.GZNUD[2018]11).

\end{document}